\documentstyle[aps,floats,manuscript,epsfig]{revtex}
\newcommand{\beq}{\begin{equation}}
\newcommand{\eeq}{\end{equation}}
\newcommand{\ba}{\begin{array}}
\newcommand{\ea}{\end{array}}
\begin{document}
\title{Exact analysis of the combined data of SNO and Super-Kamiokande}

\author{V. Berezinsky}

\address{INFN, Laboratori Nazionali del Gran Sasso, I-67010 Assergi
(AQ), Italy} 

\maketitle
\begin{abstract}
Comparison of solar-neutrino signals in SNO \cite{SNO} and 
Super-Kamiokande (SK) \cite{SK} detectors results in discovery of 
$\nu_e \rightarrow \nu_{\mu,\tau}$ oscillations at level 
$3.1 - 3.3\, \sigma$ \cite{SNO}. This comparison involves the assumption of 
neutrino spectrum and a choice for the thresholds of detection in
both experiments. In this Letter we obtain an exact formula for
the comparison of the signals which is valid for arbitrary spectra and 
thresholds. We find that the no-oscillation hypothesis  
is excluded at 3.3$\,\sigma$. If the energy-dependent component of 
the survival  probability for electron neutrinos is small as compared 
with the average value, i.e. in the case of small distortion of the 
observed spectrum,
the oscillation hypothesis can also be tested to similar accuracy.
The oscillation to sterile neutrino only, is excluded at $3.3 \sigma$
level, and oscillation to active neutrinos is confirmed at the same
level, though with some reservations.

\end{abstract}

%

Comparison of Charged Current (CC) solar-neutrino signal in SNO \cite{SNO} 
with Elastic Scattering (ES) signal in Super-Kamiokande (SK) \cite{SK} 
has revealed the 
presence of signal from $\nu_{\mu,\tau}$ neutrinos in SK detector,
which evidences for neutrino oscillation. At present this proof of
oscillation exists at level $3.1 - 3.3\, \sigma$, being limited mostly
by systematic errors and uncertainties in calculations of 
CC cross-section: $\nu_e+d \rightarrow p+p+e^-$ (for the recent
discussion of the latter and the calculations of radiative corrections
to the cross-section see Ref.\cite{Beacom}). The basic idea 
of extracting the signal from another active neutrino component 
$\nu_a=\nu_{\mu,\tau}$ comes from the fact that in CC scattering 
in SNO experiment the flux of $\nu_e$ neutrinos is measured, while 
the signal in SK is provided by all active neutrinos including $\nu_e$.    
SNO gives the total flux of $^8$B electron neutrinos, 
{\em assuming the standard spectrum}\cite{st-sp} 
as $\phi_{SNO}^{CC}=1.75 \pm 0.148 \times 10^6$~cm$^{-2}$s$^{-1}$.
Here and everywhere below we conservatively use the upper systematic 
errors and sum different errors quadratically. 
The total $^8$B neutrino flux detected by SK ({\em also assuming the
standard flux}) is 
$\phi_{SK}^{ES}=2.32 \pm 0.085 \times 10^{6}$~cm$^{-2}$s$^{-1}$.
If one assumes that only electron and sterile  neutrinos are present, 
the flux above  
is the flux of $\nu_e$ neutrinos. If to allow the presence of other
active neutrinos $\nu_a$, the flux above is $\phi_{\nu_e}+0.154 \phi_a$,
where  0.154 is the ratio of cross-sections $\nu_a e$ and $\nu_e e$. 
In any case the  3.3$\,\sigma$ excess 
$\phi_{SK}^{ES}-\phi_{SNO}^{CC}=0.57 \pm 0.17 \times 10^6$~
cm$^{-2}$s$^{-1}$ \cite{SNO} signals 
about presence of another component of active $\nu_a$ neutrinos. 
Note, that both fluxes are obtained assuming the standard (SSM) flux and using 
for their calculations the different thresholds for electron detection. 
Adjusting the thresholds, the SNO collaboration arrives at 3.1$\,\sigma$ 
flux difference $0.53\pm 0.17 \times 10^{6}$~cm$^{-2}$s$^{-1}$ \cite{SNO}. 

An interesting and effective method of comparison of the signals from 
SNO and SK was suggested in Refs.(\cite{Vill}). It is based on the 
demonstrated property that response functions for SNO and SK are
approximately equal at the appropriate choice of SNO and SK
thresholds. 

This method has been further refined in Ref.\cite{Lisi}. Choosing the
appropriate thresholds the authors verified the presence of
ocillation to active neutrinos at level $3.1\;\sigma$. Among other
interesting results the authors have obtained the value of $f_B$, 
which characterises possible deviation of boron neutrino flux from the
SSM prediction, equal to $1.03^{+0.50}_{-0.58}$.

In Ref.\cite{Giunti} it was demonstrated that in the framework of the
Unified Approach \cite{UA} the above results 
correspond to the probability of $\nu_e \rightarrow \nu_{\mu,\tau}$ 
to be larger than zero at 99.89\% CL.   

Including the results of other solar neutrino experiments and 
astrophysical information about neutrino production increase
considerably the significance of discovery of neutrino 
oscillation \cite{Bahcall}.

In this Letter we shall derive an exact and simple formula valid for
arbitrary spectra and thresholds. This formula will be obtained for an
arbitrary process of $\nu_e$ disappearance (e.g. oscillation
accompanying by the decay of a mass-eigenstate \cite{decay}),
characterised by $\nu_e$ survival probability $P_{ee}(E)$. The
oscillation, as most interesting and realistic case, will be analysed 
in detail later and anticipating it we shall keep the subscript ``osc'' 
(for oscillation) as a notation from the beginning.

The basic quantity that we shall analyse is the ratio of the electron rate in 
SK, $R_{\nu_a}^{SK}$, produced by active neutrinos $\nu_a=\nu_{\mu,\tau}$ 
and the total electron rate $R_{\rm tot}^{SK}$:
\beq
r_{\rm osc}= R_{\nu_a}^{SK}/R_{\rm tot}^{SK}.
\label{ratio}
\eeq
Let us  introduce the following definitions.

1. The exact flux of $\nu_e$ neutrinos reaching a detector is defined as 
\beq
\Phi_{\nu_e}(E_\nu)=\Phi_B\phi_{SSM}(E_{\nu})P_{ee}(E_{\nu}),
\label{eflux}
\eeq
where $\Phi_B$ is the total boron neutrino flux at production,
the value of which is not specified, and  $\phi_{SSM}(E_{\nu})$ 
is the spectrum at production, normalised as 
\beq
\int \phi_{SSM}(E_{\nu})dE_{\nu}=1.
\label{SSMnorm}
\eeq

2. According to the definition of SK collaboration, the detected 
$^8$B neutrino flux, $\phi_{SK}^{ES}$, determines the total rate in SK above
threshold $T_{th}^{SK}$,
{\em if the neutrino spectrum in detector is standard}:
\beq
R_{\rm tot}^{SK}=\phi_{SK}^{ES}\int_{T_{th}^{SK}}^{T^{max}}dT
\int dT^{\prime}R_{SK}(T,T^{\prime})
\int_{E_{\nu}^{min}(T^{\prime})}^{E_{\nu}^{max}(T^{\prime})}dE_{\nu}%
\phi_{SSM}(E_{\nu})\sigma_{\nu_e e}(E_{\nu},T^{\prime}),
\label{SKtotrate}
\eeq
where $T^{\prime}$ and $T$ are the real and the measured kinetic energies of
the electron, respectively,  and $R_{SK}(T,T^{\prime})$ is
energy resolution function of the Super-Kamiokande detector \cite{SK1}. 
$\phi_{SK}^{ES}$ can be interpreted as the flux of $\nu_e$ neutrinos in
case of $\nu_e \rightarrow \nu_s$ conversion, and as 
$\phi_{SK}^{ES}=\phi_{\nu_e}+0.154\phi_{\nu_a}$ in case of 
$\nu_e \rightarrow \nu_s$ conversion, where 0.154 is a ratio 
$\sigma(\nu_a e)/\sigma(\nu_e e)$.

3. The flux of electron neutrinos from $^8$B decays,
$\phi_{SNO}^{CC}$,  
is determined by SNO collaboration from the CC-reaction rate above 
threshold $T_{\rm th}^{SNO}$, {\em assuming the standard spectrum}:
\beq
R_{SNO}^{CC}=\phi_{SNO}^{CC}\int_{T_{\rm th}^{SNO}}^{T_{max}}dT%
\int dT^{\prime}R_{SNO}(T,T^{\prime})
\int_{E_{\nu}^{min}(T^{\prime})}^{E_{\nu}^{max}(T^{\prime})}dE_{\nu}%
\phi_{SSM}(E_{\nu})\sigma_{CC}(E_{\nu},T^{\prime}),
\label{CCrate1}
\eeq
where $R_{SNO}(T,T^{\prime})$ is energy resolution function of SNO
\cite{SNO}.

4. On the other hand the same rate $R_{SNO}^{CC}$ can be expressed via
the exact flux $\Phi_{\nu_e}(E_\nu)$, as given by Eq.(\ref{eflux}): 
\beq
R_{SNO}^{CC}=\Phi_B\int_{T_{\rm th}^{SNO}}^{T_{max}}dT 
\int dT^{\prime}R_{SNO}(T,T^{\prime})
\int_{E_{\nu}^{min}(T^{\prime})}^{E_{\nu}^{max}(T^{\prime})}dE_{\nu}
\phi_{SSM}(E_{\nu})P_{ee}(E_{\nu})\sigma_{CC}(E_{\nu},T^{\prime}).
\label{CCrate2}
\eeq

5. Finally, we define two quantities,
$J_i$ and $J_i\{P_{ee}\}$ 
\begin{eqnarray}
J_i\{P_{ee}\}& \equiv &\int_{T_{\rm th}^{i}}^{T_{max}}dT 
\int dT^{\prime}R_{i}(T,T^{\prime})%
\int_{E_{\nu}^{min}(T^{\prime})}^{E_{\nu}^{max}(T^{\prime})}dE_{\nu}%
\phi_{SSM}(E_{\nu})P_{ee}(E_{\nu})\sigma_{i}(E_{\nu},T^{\prime}),
\label{JP}\\
J_i &\equiv & J_i\{P_{ee}=1\}, \label{J}
\end{eqnarray}
for $i=SNO,~SK,~ ~\sigma_{i}=\sigma_{CC}$ for SNO and 
$\sigma_{i}=\sigma_{\nu_e e}$ for SK.  
The ratio $J_i\{P_{ee}\}/J_i$ determines the average 
survival probability $\langle P_{ee} \rangle_i$ for $i=$SNO, SK. This
ratio is $P_{ee}$ in case of constant survival probability. 

We are now ready to obtain the exact expression for the ratio 
$r_{\rm osc}$ from Eq.(\ref{ratio}). Rearranging it as
$r_{\rm osc}=1-R_{\nu_e}^{SK}/R_{\rm tot}^{SK}$, and using for SK rate induced
by $\nu_e$ ~ $R_{\nu_e}^{SK}=\Phi_BJ_{SK}\{P_{ee}\}$ we obtain 
\beq
r_{\rm osc}=1-\frac{\Phi_B J_{SK}\{P_{ee}\}}{\phi_{SK}^{ES}J_{SK}},
\label{r}
\eeq
where $\Phi_B$ is given by Eqs.(\ref{CCrate1}) and (\ref{CCrate2}) as
\beq
\Phi_B=\phi_{SNO}^{CC}\frac{J_{SNO}}{J_{SNO}\{P_{ee}\}}.
\label{PhiB}
\eeq

We can decompose $P_{ee}(E)$ in the energy
independent part $\bar{P}_{ee}$, e.g. the average value in one of 
the experiments, and a small (according to experimental data) energy 
dependent part $\delta P(E)$:~~ $P_{ee}(E)=\bar{P}_{ee}+ \delta P(E)$. 
Then  we can re-write Eq.(\ref{r}) as 
\beq
r_{\rm osc}=1-\frac{\phi_{SNO}^{CC}}{\phi_{SK}^{ES}}%
\frac{1+ J_{SK}\{\frac{\delta P_{ee}}{\bar{P}_{ee}}\}/J_{SK}}%
{1+ J_{SNO}\{\frac{\delta P_{ee}}{\bar{P}_{ee}}\}/J_{SNO}}=
1-\frac{\phi_{SNO}^{CC}}{\phi_{SK}^{ES}}
\frac{1+\langle\frac{\delta P_{ee}}{\bar{P}_{ee}}\rangle_{SK}}
{1+\langle\frac{\delta P_{ee}}{\bar{P}_{ee}}\rangle_{SNO}},
\label{osc}
\eeq
where the angle brackets mean energy averaging. The formula (\ref{osc})
is exact. It is valid in particular even when $\delta P(E)$ is not small 
in comparison with $\bar{P}_{ee}$. It is derived for any process of 
$\nu_e$ disappearance, including the oscillation. 

The correction factor in Eq.(\ref{osc}),
\beq
K_{\rm corr}=
\frac{1+\langle\frac{\delta P_{ee}}{\bar{P}_{ee}}\rangle_{SK}}
{1+\langle\frac{\delta P_{ee}}{\bar{P}_{ee}}\rangle_{SNO}},
\label{corr}
\eeq
is close to 1 in each of two cases: when 
{\it (i)} $\delta P_{ee}/\bar{P}_{ee}$ is small or when 
{\it (ii)} average values $\langle\delta P_{ee}/\bar{P}_{ee}\rangle$
for SNO and SK are close to each other. As we will see in the 
physically relevant cases both conditions work simultaneously. 

The first condition is satisfied because the observed distortion of
spectrum is small. The correction factor (\ref{corr}) is invariant
relative to decomposition $P_{ee}(E)=\bar{P}_{ee}+\delta P_{ee}(E)$ 
with arbitrary $\bar{P}_{ee}$. Since the distortion of spectrum is small 
we always can choose the value of  $\bar{P}_{ee}$ such  that 
$\delta P_{ee}(E)$ changes its sign in the middle of the observed
energy interval.  It provides the smallness of 
$\langle\delta P_{ee}/\bar{P}_{ee}\rangle_i$. The condition {\it (ii)}
will further diminish $1-K_{\rm corr}$.

In general case $K_{\rm corr}$ can be obtained directly from
observational data using $\delta P_{ee}(E)$ for each energy bin.

We shall consider now  the application of Eq.(\ref{osc}) for neutrino 
oscillations assuming that {\it oscillation is the only way of 
$\nu_e$ disappearance}.  

To test the different hypotheses we shall treat $r_{\rm osc}$, given by 
Eq.(\ref{osc}), as experimental value to be compared with a
theoretical value $r_{\rm osc}=0$.
The latter is the predicted value for the hypothesis of oscillation to
sterile neutrino only, as in the case of two neutrino mixing (hereafter
we shall refer to this case as {\em sterile-neutrino oscillation}). 
In the case of oscillation to active-neutrinos $\nu_a$, to be referred
to as {\it active-neutrino oscillation},  $r_{\rm osc}=0$ corresponds
to absence of $\nu_e \rightarrow \nu_a$ oscillation. 

For the case $P_{ee}(E)=const$ we obtain 
\beq
r_{\rm osc}= 1-\frac{\phi_{SNO}^{CC}}{\phi_{SK}^{ES}}=0.246 \pm 0.0694.
\label{no-osc}
\eeq
In particular, this value of the ratio is exact for
no-oscillation case 
$P_{ee}=1$, and formally it is 3.54$\,\sigma$ away from no-oscillation value
$r_{\rm osc}=0$.

However, evaluation of the error for the ratio 
$\phi_{SNO}^{CC}/\phi_{SK}^{ES}$ does not correspond to to the usual
connection with the confidence level, because the distribution is not
Gaussian. More correct evaluation of the error is as follows.

In the plane $(\phi_{SNO}^{CC}\,\,,\,\,\phi_{SK}^{ES})$ one plots 
experimentally measured point and $1\,\sigma,~2\,\sigma$ and $3\,\sigma$
contours around it (see Fig.~1).  One can see that the contour $3.33\,\sigma$ 
touches the line $r_{\rm osc}=0$. 
\vspace{-5mm}
\begin{figure}[hp]
\begin{center}
\epsfig{bbllx=75pt,bblly=250pt,bburx=530pt,bbury=670pt,%
file=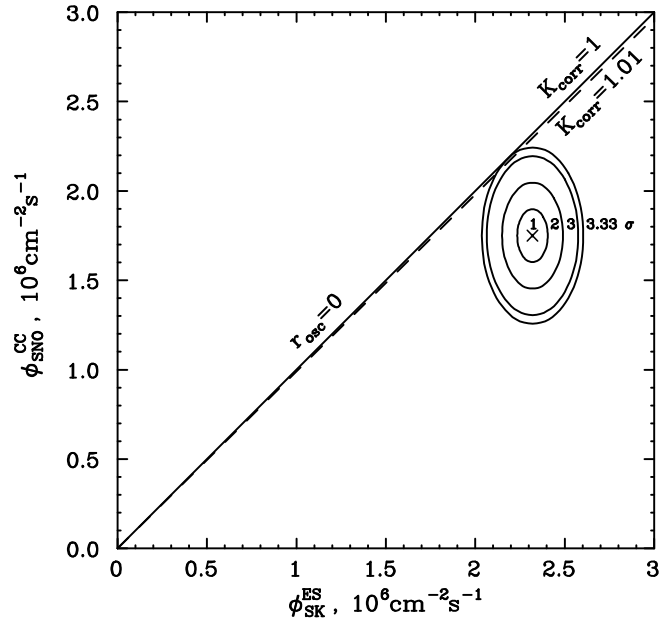,height=9cm}
\end{center}
\vspace{-5mm}
\caption[a]{ SNO and Super-Kamiokande fluxes allowed at $1\;\sigma,~~
2\;\sigma$ and $3.33\;\sigma$ levels in comparison with no-oscillation 
value $r_{\rm osc}=0$ for two values of correction factor 
$K_{\rm corr}=1$ and $K_{\rm corr}=1.01$. The experimental point is 
$3.3\;\sigma$ away from $r_{\rm osc}=0$ which corresponds to absence 
of active neutrino oscillation.
\label{fig1}
} 
\end{figure}

In the general case to
obtain the value $r_{\rm osc}$ from Eq.(\ref{osc}) we evaluate
$K_{\rm corr}$ in the following way.

First we assume, and later verify, that correction factor for a
tested hypothesis is given by value $K_{\rm corr} \sim 1$
within very narrow interval $\Delta K_{\rm corr}$, so that we can
treat $K_{\rm corr}$ just as a number. Then using 
$\phi_{SNO}^{CC}/\phi_{SK}^{ES}=0.754 \pm 0.0694$ we find formally for this 
hypothesis 
\beq
r_{\rm osc}=1-0.754\, K_{\rm corr} \pm 0.0694 K_{\rm corr}
\label{roscpee} 
\eeq

To obtain the range of allowed values of $K_{\rm corr}$ one must compute the
correction factors for all points in the parameter oscillation space
allowed by all available data e.g. at $1 \sigma$, with the Super-Kamiokande 
spectra data being the most important ingredient. 

If the starting assumption about narrow interval $\Delta K_{\rm corr}$
is confirmed, one can proceed further choosing the maximum value of 
$K_{\rm corr}$ from obtained range and inserting it to Eq.(\ref{roscpee}).
$K_{\rm corr}^{\rm max}$ gives the smallest value of $r_{\rm osc}$
with the largest error, which we conservatively take as the final
result.

In both cases, sterile-neutrino oscillation and active-neutrino
oscillation, a statistically significant deflection from 
$r_{\rm osc}=0$ is a proof of the hypothesis (absence of 
sterile-neutrino oscillation, and presence of active-neutrino
oscillation), though for sterile neutrinos this criterion is formally
more strict, because $r_{\rm osc}=0$ is the theoretical prediction of 
sterile-neutrino hypothesis. 

For sterile neutrinos $K_{\rm corr}$ has been calculated \footnote{The 
analysis of
sterile-neutrino oscillation, including the calculations of 
$K_{\rm corr}$ has been 
performed by M.C.Gonzalez-Garcia and C.Pena-Garay. They have also
calculated the data of Table 1.} 
for all points 
in the sterile oscillation parameter space $(\Delta m^2, \tan^2\theta)$
allowed by  
the Super-Kamiokande event rates and the day-night spectra at  $1 \sigma$.
It results in $0.97\leq K_{\rm corr}\leq 1.03$,
which, as expected, does not differ much from the energy-constant value 1.
Using  Eq.(\ref{roscpee}) with $K_{\rm corr}^{\rm max}= 1.03$
one obtains that the sterile oscillation is excluded  with formal
significance  $\geq 3.1\,\sigma$. For the realistic evaluation significance
is less (see Fig.1).

If to add to the Super-Kamiokande data all other pre-SNO data 
(including the measured rates in the chlorine and gallium experiments),
the status of sterile-neutrino oscillation becomes  further disfavoured.
In this case the range of allowed values of $K_{\rm corr}$ 
is slightly shifted to lower
values $0.93\leq K_{\rm corr}\leq 1$. With this range, the
sterile-neutrino  oscillation
is excluded at significance better than $3.3\,\sigma$, as it can be
seen from Fig.1.

For active neutrino-oscillation the analysis can be made in identical
way.  In this case the correction factors have to be calculated for
all points in the parameter space $(\Delta m^2, \tan^2\theta)$ allowed 
by all experimental data. In fact we used the
range of calculated values of $K_{\rm corr}$  for the points
of best fit solutions \cite{GPS} only (see Table 1) as 
$0.944 \leq K_{\rm corr}\leq 1.002$.
From Eq.(\ref{roscpee}) we obtain formally $r_{\rm osc}=0.244 \pm 0.0695$.
\begin{table}[ht]
\caption[]{Correction factors $K_{\rm corr}$ for the best fit solutions}
\begin{eqnarray}
\begin{array}{lccccr}
{\rm solution}   &\tan^2\theta    &  \Delta m^2,~ {\rm eV}^2  & K_{\rm  corr}\\
\hline
{\rm LMA}        &   0.365        &  3.65\cdot 10^{-5}    &   1.002\\ 
{\rm SMA}        &   0.00061      &  5.01\cdot 10^{-6}    &   0.944 \\
{\rm LOW}        &   0.708        &  1.03\cdot 10^{-7}    &   0.993 \\
{\rm JS}^2       &   1.000        &  5.46\cdot 10^{-11}   &    0.995\nonumber        \\
\end{array}
\end{eqnarray}
\end{table}
\vspace{-2mm}
From Fig.1 we see that that observed fluxes for this range of 
correction factors are $3.3 \sigma$ away 
from $r_{\rm osc}=0$, which corresponds to 
hypothesis about absence of active-neutrino oscillation. 

In conclusion, we have derived the exact formula (\ref{osc}) for
calculation of ratio of the rate in Super-Kamiokande from $\nu_{\mu,\tau}$ 
neutrinos to the total rate. This ratio, $r_{\rm osc}$, characterises the
oscillation of electron neutrino to another active neutrino. The
formula is valid for arbitrary thresholds in SNO and SK experiments, 
and arbitrary neutrino spectra. In particular,   
smallness of $\delta P_{ee}/\bar{P}_{ee}$ is not
required for validity of Eq.(\ref{osc}). 
We have shown that  this ratio can be used to test the hypotheses
of no-oscillations and oscillations to sterile and active neutrinos.

For the no-oscillation case  
the ratio $r_{\rm osc}$ is given by Eq.(\ref{no-osc})
and this case is excluded at 3.3$\,\sigma$ by comparison with 
theoretical no-oscillation value $r_{\rm osc}=0$.  

 For a small
energy-dependent component of survival probability,  
$\delta P_{ee}/\bar{P}_{ee}$, the ratio is described with a good 
accuracy by the same  
Eq.(\ref{no-osc}), which would exclude $\nu_e \rightarrow \nu_s$
oscillation at 3.3$\sigma$. This hypothesis is further tested
evaluating numerically the correction factor $K_{\rm corr}$ for
all points in the parameter space $(\Delta m^2, \tan^2\theta)$
in the pre-SNO allowed region for 
sterile-neutrino  oscillations. 
It is found that sterile-neutrino oscillation is excluded at  
3.3$\sigma$.

For oscillation to active neutrinos our analysis is not as accurate as
for sterile neutrinos. The correction factors are calculated only for 
the best-fit points in the parameter space.  With these data the
oscillations to active
neutrino is confirmed at 3.3$\sigma$. 

With statistical error reduced by factor 2, the
the significance of above results will reach 3.5$\sigma$.

\section*{Acknowledgement}
I am deeply grateful to M.C.Gonzalez-Garcia and C.Pena-Garay for many
detailed discussions and for numerical calculations. I thank also
S.Oser for important remarks and M.Lissia for valuable discussions and  
help with building the graph.

\end{document}